\documentclass[number,preprint,review,12pt]{elsarticle}
\biboptions{sort&compress}
\usepackage{lineno}
\usepackage[pdftex,plainpages=false,pdfpagelabels,colorlinks=true,citecolor=Blue,linkcolor=DarkBlue,urlcolor=DarkRed,filecolor=green,bookmarksopen=true]{hyperref}
\modulolinenumbers[5]

\def\txtcolorSec{black}

\usepackage[utf8]{inputenc}
\usepackage[top=1in,bottom=1in,left=1in,right=1in]{geometry}
\usepackage{amsmath,amssymb,amsfonts,amsthm,accents,mathrsfs,bm}
\usepackage{graphicx,subcaption}
\usepackage{wrapfig,epsfig}
\usepackage{setspace}         
\usepackage{comment}
\usepackage{enumerate,makeidx}
\usepackage{dcolumn}          
\usepackage{booktabs,colortbl,array}
\usepackage{sectsty,lastpage}
\usepackage[displaymath,textmath,sections,graphics,floats]{preview}
\PreviewEnvironment{enumerate,tabular}
\usepackage{hyphenat}
\usepackage{nicematrix}
\usepackage{soul}
\usepackage{algorithm, algpseudocode}
\DeclareCaptionLabelFormat{myformat}{\textcolor{\txtcolorSec}{Code block #2}}
\usepackage{multicol,multirow} 
\usepackage{xstring}
\newcommand{\textuser}[2]{%
    \IfEqCase{#1}{
        {1}{\textit{#2}}
        {2}{\texttt{#2}}
        {3}{\textnormal{#2}}
    }[\PackageError{textuser}{Undefined option to textuser: #1}{}]%
}%
\usepackage{csquotes}
\usepackage{makecell}
\usepackage{caption}
\journal{Journal of Materials Research}

\biboptions{sort&compress}

\begin{document}
\begin{frontmatter}
\title{Geometry-Dependent Crack Interaction and Toughening in Graphene}

\author[mymainaddress0]{Suyeong Jin}
\author[mymainaddress1]{Jung-Wuk Hong}
\author[mymainaddress3]{Alexandre F. Fonseca\corref{mycorrespondingauthor1}}
\cortext[mycorrespondingauthor1]{Corresponding author}\ead{afonseca@ifi.unicamp.br}

\address[mymainaddress0]{Department of Mechanical Engineering, Pukyong National University, 45 Yongso-ro, Nam-gu, Busan 48513, Republic of Korea}
\address[mymainaddress1]{Department of Civil and Environmental Engineering, Korea Advanced Institute of Science and Technology, 291 Daehak-ro, Yuseong-gu, Daejeon 34141, Republic of Korea}
\address[mymainaddress3]{Universidade Estadual de Campinas (UNICAMP), Instituto de F\'{i}sica Gleb Wataghin, Departamento de F\'{i}sica Aplicada, 13083-859, Campinas, SP, Brazil}

\begin{abstract}
The interaction between neighboring cracks has been shown to strongly
influence the fracture behavior of graphene. While previous studies
focused primarily on crack spacing, the role of crack width remains
poorly understood. Here, computational simulations are performed to
investigate the coupled effects of crack width and inter-crack spacing
$(W_\text{gap})$ on the tensile response of graphene containing
parallel cracks. The results show that increasing crack width
amplifies the sensitivity of mechanical properties to crack spacing,
leading to significant enhancement of peak stress, fracture strain,
and toughness at sufficiently large $W_\text{gap}$. For narrow cracks,
crack coalescence dominates and causes brittle failure. In contrast,
wider cracks promote delayed ligament rupture, increased energy
absorption and ductile-like fracture behavior. The normalized
toughness and fracture strain exceed those of equivalent single-crack
systems by more than twofold. A crack-geometry design map is proposed
to identify regimes of crack coalescence, independent propagation, and
enhanced toughness.
\end{abstract}

\end{frontmatter}

\section{Introduction}

Graphene has attracted significant attention due to the exceptional
mechanical, electrical, and thermal properties, making it a promising
material for applications ranging from flexible electronics to
structural
nanodevices~\cite{quesnel2015graphene,han2017MSER,dhinakaran2020review}.
In particular, the high Young’s modulus and intrinsic tensile strength
have established graphene as a benchmark two-dimensional material for
studying nanoscale mechanics and fracture behavior. Despite these
outstanding properties, the practical implementation of graphene
remains strongly affected by the presence of structural defects
introduced during synthesis, transfer, and processing
procedures~\cite{Lin2009NatComm,Boggild2023NatComm,Pham2024}.
Imperfections such as vacancies, tears, holes, and cracks are commonly
observed in experimentally fabricated graphene sheets and can
substantially modify their mechanical response under external
loading. Since fracture initiation and crack propagation govern the
structural reliability of graphene-based systems, understanding how
defects influence stress redistribution and failure mechanisms is
essential for the design of mechanically robust graphene materials and
devices~\cite{Mahesh2022RSCAdv,Dongbo2024EFM}.

Extensive theoretical, experimental, and computational studies have investigated the fracture behavior of graphene containing structural defects. Previous works have examined the effects of vacancies, grain boundaries, nanoholes, and preexisting cracks on the mechanical properties and fracture toughness of graphene~\cite{Hu2015JAP,DewaCarbon2017,DewaEngFracMech2018,DewaCMS2018,YaoEngFracMech2019,Chen2020ACSANM,BrodnikPRL2021,felixPCCP2022}. Molecular dynamics (MD) simulations have shown that fracture in graphene strongly depends on crack orientation, loading direction, temperature, and strain rate, with distinct propagation characteristics observed along armchair and zigzag directions~\cite{John2014Carbon,Fujihara2015ACSNano}. Defect geometry can substantially alter local stress concentration and crack-tip propagation, leading to anisotropic fracture responses and variations in strength and toughness~\cite{Zandiatashbar2014NatComm,Liu2025JPCS}. 
Therefore, the mechanical behavior of graphene is governed by defects, geometry, spatial arrangement, and mutual interactions.

Among the various defect configurations explored in graphene,
structures containing multiple interacting cracks have attracted
growing interest. Crack interaction can modify fracture behavior
beyond that predicted for isolated
defects~\cite{Zhang2012,Lopez-Polin2015,Meng2015,Liu2025JPCS}. The
stress fields generated by neighboring cracks may overlap, resulting
in crack shielding, delayed crack propagation, or crack coalescence
depending on the crack geometry and spacing~\cite{suyeong2026IJMS}.
In particular, parallel crack configurations provide a model system
for investigating how geometric parameters control stress
redistribution and failure evolution in two-dimensional
materials. Varying the inter-crack spacing in graphene with parallel
cracks induces a transition from crack coalescence to independent
crack propagation accompanied by ductile-like deformation of the
ligament between cracks. These results demonstrated the role of crack
interaction in governing fracture mechanisms and suggested that defect
geometry can potentially be utilized as a design parameter for
tailoring the mechanical performance of graphene.

Despite the growing understanding of crack interaction in graphene,
the coupled influence of crack width and inter-crack spacing on crack
coalescence in graphene remains largely unexplored. Previous studies
primarily focused on the effects of crack length, crack orientation,
or crack spacing, while the role of crack width in controlling stress
redistribution, ligament deformation, and energy dissipation has
received comparatively little attention. Crack width can substantially
modify the local geometry of the ligament between neighboring cracks,
potentially altering both the intensity of crack interaction and the
sequence of fracture events. In particular, the interplay between
crack width and crack spacing may determine whether the ligament
undergoes rapid crack coalescence or progressive deformation prior to
rupture.

In this study, molecular dynamics simulations based on the reactive
force field (ReaxFF)~\cite{vanDuin2001} are performed to investigate
the coupled effects of crack width ($2b$) and inter-crack spacing
($W_\text{gap}$) on the fracture behavior of graphene containing
parallel cracks. Armchair and zigzag graphene structures subjected to
uniaxial tensile loading are analyzed for varying crack
geometries. The mechanical response is characterized through
stress–strain relations, peak stress, fracture strain, and toughness,
while atomistic snapshots and local stress distributions are employed
to elucidate the underlying fracture mechanisms.

The remainder of this paper is organized as follows.
Section~\ref{sec3} presents the mechanical response of graphene with
parallel cracks for different crack widths and inter-crack spacings.
The main conclusions for geometry-controlled fracture engineering in
graphene are summarized in section~\ref{sec4}. Finally,
section~\ref{sec2} describes the graphene models, crack geometries,
and molecular dynamics simulation procedures employed in this study.


\section{Results and Discussion}\label{sec3}
\begin{figure}
    \centering
    \includegraphics[width=\linewidth]{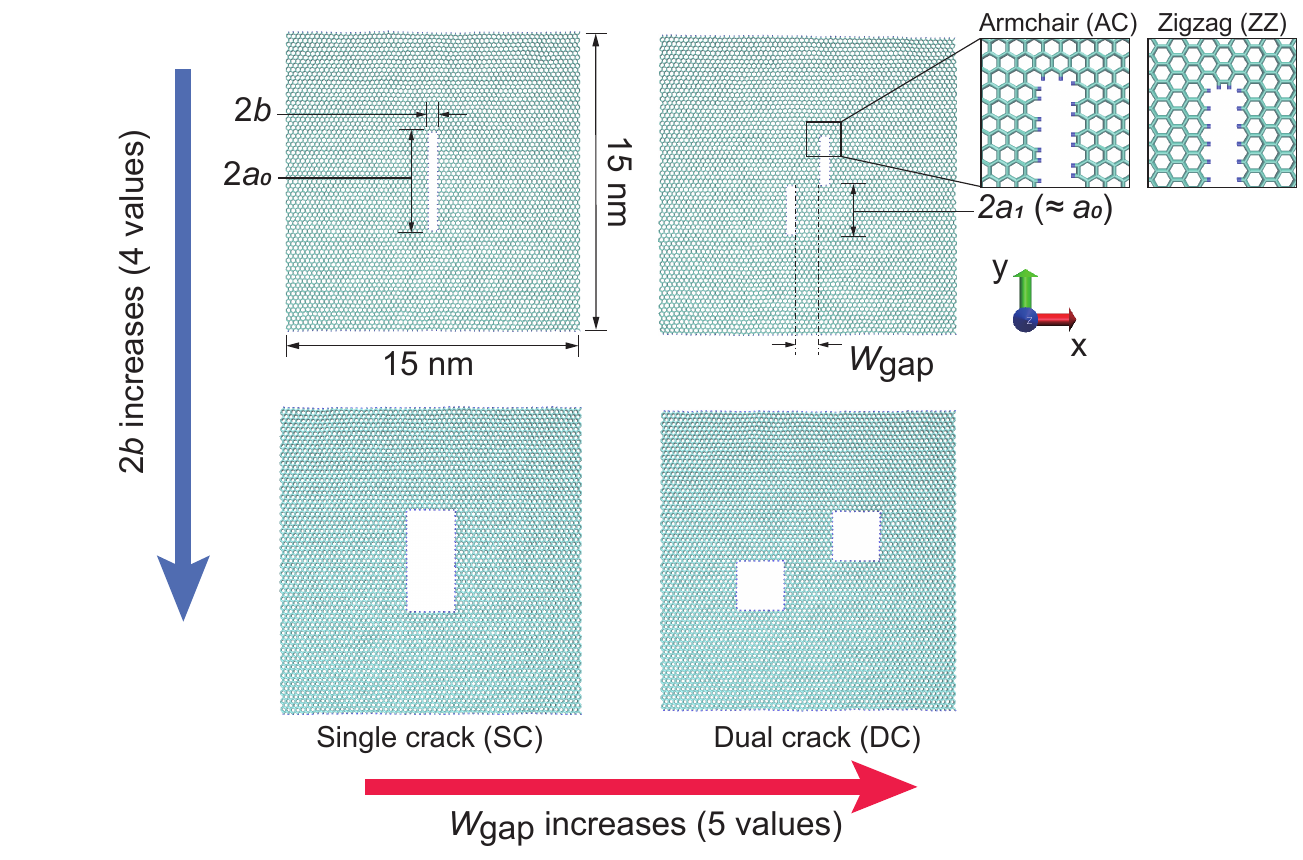}
    \caption{\label{fig:geometry}Geometry of graphene with varying
      crack width $2b$ and crack gap $W_\text{gap}$; total four single
      crack (SC) cases and 16 cases of dual crack (DC) are prepared
      for armchair (AC) and zigzag (ZZ) chirality, respectively. The
      single crack of length is 2$a_0$ and each crack length of
      parallel cracks, 2$a_1$, where $2a_1\approx a_0$, is separated
      by $W_\text{gap}$.  A magnified view of the atomic structure
      surrounding the cracks is presented, where carbon and hydrogen
      atoms are depicted in cyan and blue, respectively.}%
\end{figure}

\begin{figure}[tbp]
    \centering
    \includegraphics[width=\linewidth]{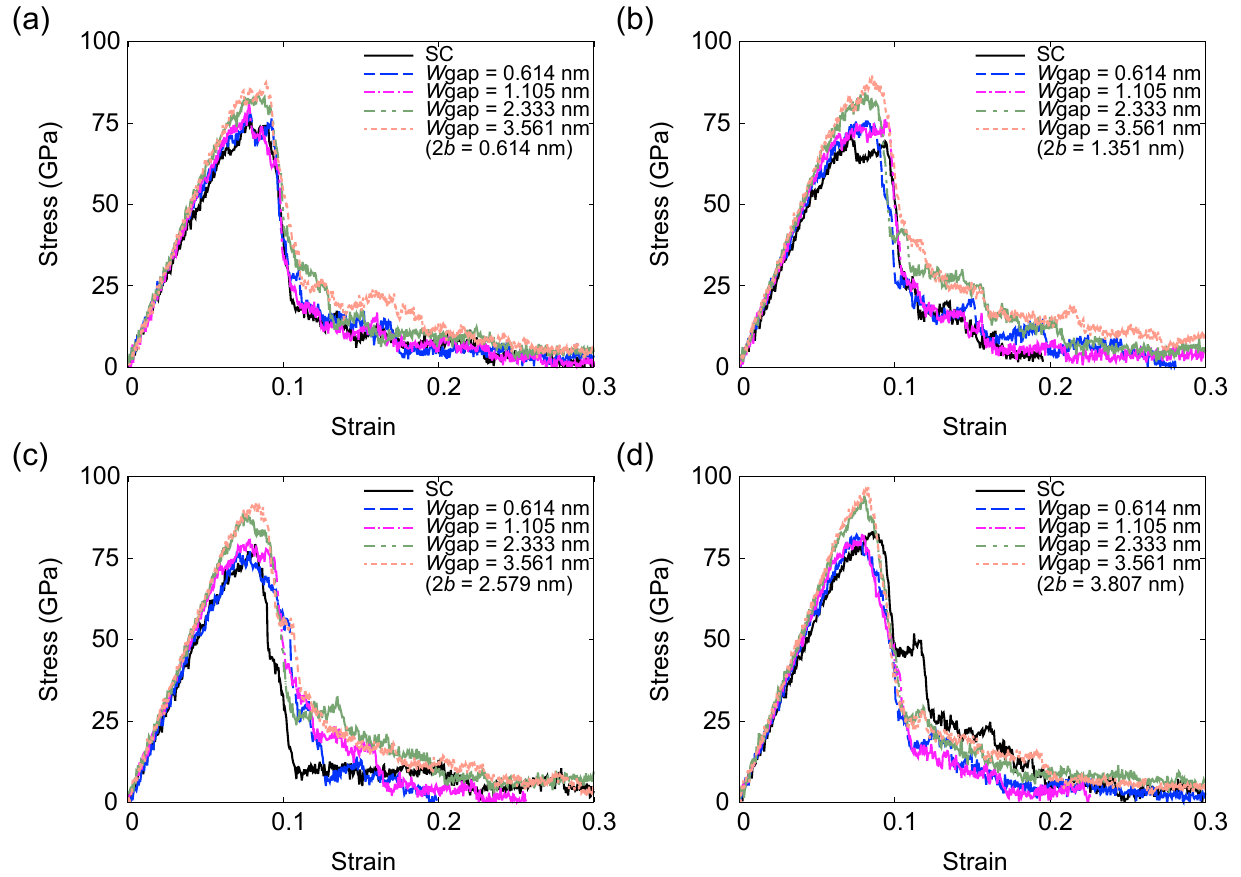}
    \caption{\label{fig:ss-curve-AC}Stress-strain curves of armchair (AC) graphene structure with dual cracks under uniaxial tensile loading along the \textit{x}-direction, simulated using the ReaxFF interatomic potential. Results are grouped by crack width (a) $2b=0.614$~nm, (b) $2b=1.351$~nm, (c) $2b=2.579$~nm, and (d) $2b=3.807$~nm. Each panel presents four dual-crack configurations with increasing inter-crack gap $W_\text{gap}$ alongside the corresponding single-crack (SC) reference.} 
\end{figure}

\begin{figure}[tbp]
    \centering
    \includegraphics[width=\linewidth]{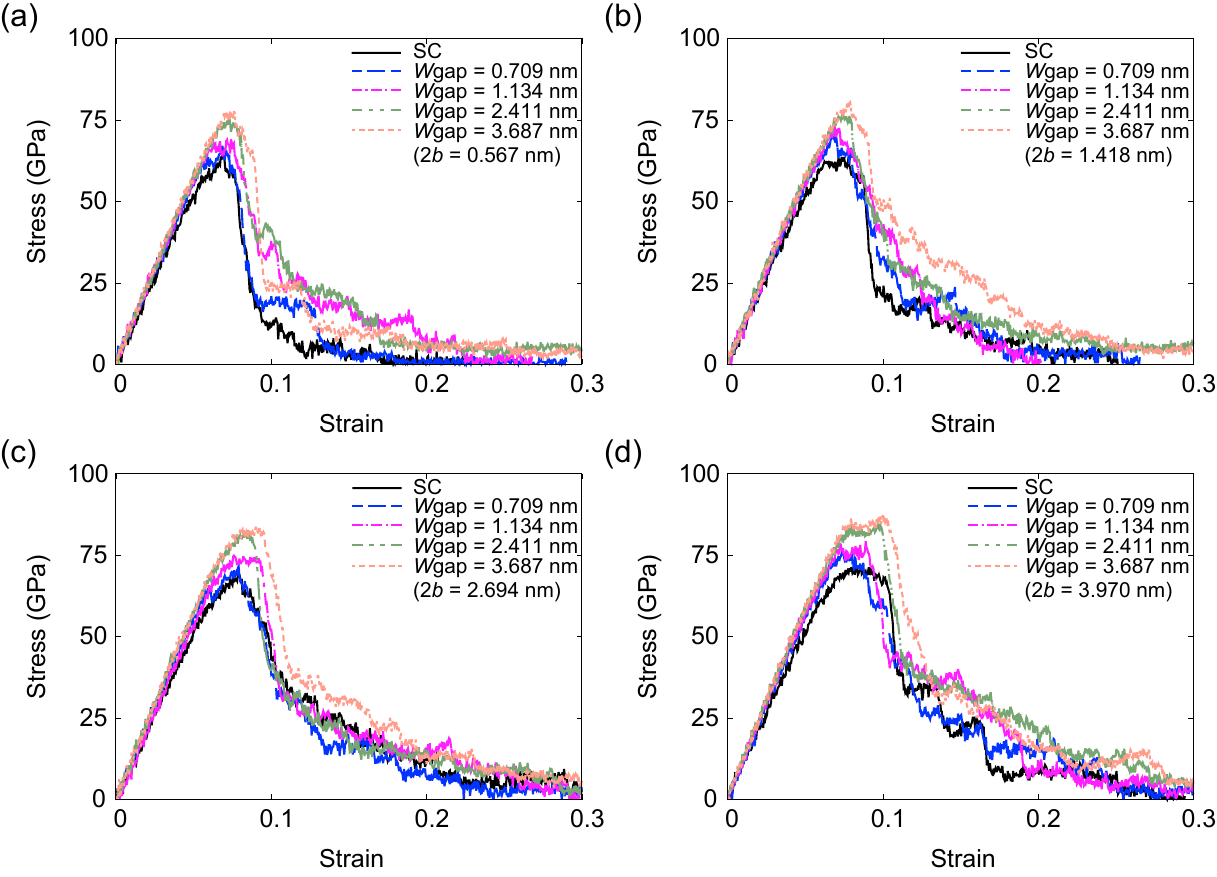}
    \caption{\label{fig:ss-curve-ZZ}Stress-strain curves of zigzag (ZZ) graphene structure with dual cracks under uniaxial tensile loading along the \textit{x}-direction, simulated using the ReaxFF interatomic potential. Results are grouped by crack width (a) $2b=0.567$~nm, (b) $2b=1.418$~nm, (c) $2b=2.694$~nm, and (d) $2b=3.970$~nm. Each panel presents four dual-crack configurations with increasing inter-crack gap $W_\text{gap}$ alongside the corresponding single-crack (SC) reference.} 
\end{figure}

The graphene structures and crack geometries considered in the present
study are illustrated in Fig.~\ref{fig:geometry}.  Section~\ref{sec2}
describes the meaning of the abbreviations and geometric quantities.
Figs.~\ref{fig:ss-curve-AC} and \ref{fig:ss-curve-ZZ} show the stress--strain curves of armchair (AC) and zigzag (ZZ) graphene containing single-crack (SC) and dual-crack (DC) configurations with different crack widths ($2b$) and inter-crack gaps ($W_\text{gap}$), respectively. All configurations exhibit an initial approximately linear elastic regime, followed by nonlinear deformation and stress reduction associated with crack propagation and fracture.
For both AC and ZZ graphene, the stress--strain curves of the DC configurations generally shifted upward as $W_\text{gap}$ increased within each crack-width group. This trend indicates that increasing the spacing between neighboring cracks modifies the tensile response of graphene with parallel cracks. The upward shift is observed more clearly in several larger crack-width cases, suggesting that the influence of crack spacing becomes more pronounced when the crack width increases.
Post-peak softening was also dependent on the crack geometry. In several DC configurations, the stress decreased more gradually and persisted over a broader strain range than in the corresponding SC cases, suggesting that crack interaction alters the fracture progression after peak loading. 
The stress--strain responses in Figs.~\ref{fig:ss-curve-AC} and \ref{fig:ss-curve-ZZ} are further quantified below in terms of toughness. The peak stress and fracture strain are quantified in \ref{Appendix:data}.


\begin{figure}[tbp]
    \centering
    \includegraphics[width=\linewidth]{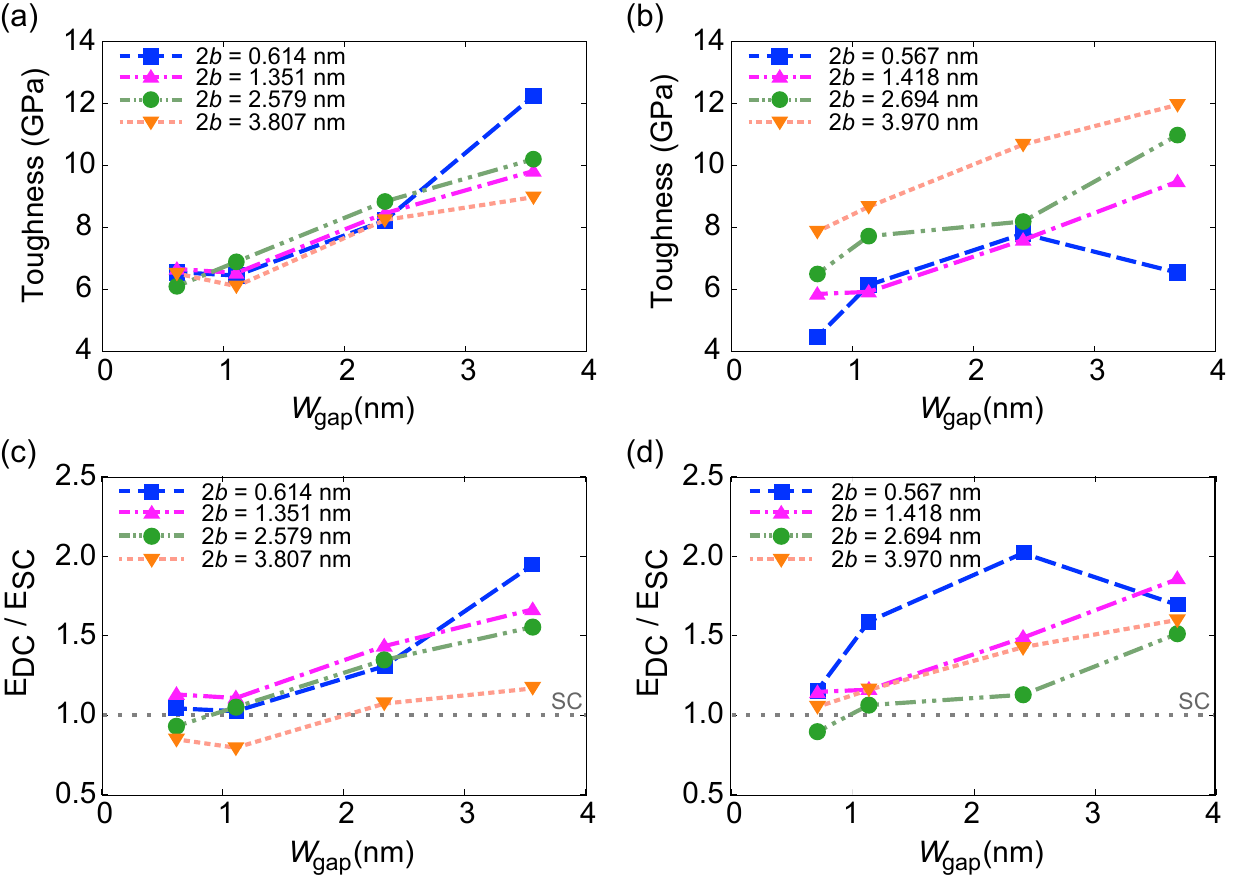}
    \caption{\label{fig:toughness-vs-Wgap}Toughness of dual-crack (DC)
      graphene sheets as a function of crack gap
      $W_\text{gap}$. Panels show results for (a) armchair (AC) and
      (b) zigzag (ZZ) chiralities. Panels (c) and (d) show the
      corresponding toughness ratio $E_\text{DC}/E_\text{SC}$ relative
      to the single-crack (SC) reference with the same crack width
      $2b$, for AC and ZZ, respectively. The horizontal dashed line at
      unity in (c) and (d) indicates the SC reference level. Toughness
      is defined as the area under the stress–strain curve up to the
      fracture strain $\varepsilon_f$.}
\end{figure}

Fig.~\ref{fig:toughness-vs-Wgap} represents the toughness of DC
graphene as a function of $W_\text{gap}$. For both chiralities,
toughness generally increased with $W_\text{gap}$, indicating enhanced
energy absorption at larger $W\text{gap}$. The largest enhancement is
observed for narrow-crack AC graphene, where the toughness reaches
12.25 GJ/m$^3$ at $W_\text{gap}=3.56$~nm, compared with 6.29 GJ/m$^3$
for the corresponding SC configuration.  In contrast, the narrow-crack
ZZ case shows a non-monotonic response, increasing to 7.80 GJ/m$^3$ at
$W_\text{gap}=2.41$~nm before decreasing to 6.54 GJ/m$^3$ at
$W_\text{gap}=3.69$~nm, consistent with the fracture-strain trend in
\ref{Appendix:data}. The normalized toughness ratio,
$E_\text{DC}/E_\text{SC}$, ranges from 0.79 to 2.02, much broader than
the normalized peak-stress range of 0.98--1.26 (see
\ref{Appendix:data}). At small $W_\text{gap}$, several wide-crack
configurations exhibited ratios below unity, while at large
$W_\text{gap}$, all crack-width groups exceeded unity. A
chirality-dependent crossover is also observed: AC graphene shows
higher toughness for narrow cracks, while ZZ graphene surpasses AC
graphene for wide cracks at large $W_\text{gap}$. Overall, these
results indicate that crack interaction affects energy absorption more
strongly than peak strength and promotes a transition from rapid crack
coalescence at small $W_\text{gap}$ to more progressive fracture at
large $W_\text{gap}$.

\begin{figure}[tbp]
    \centering
    \includegraphics[width=\linewidth]{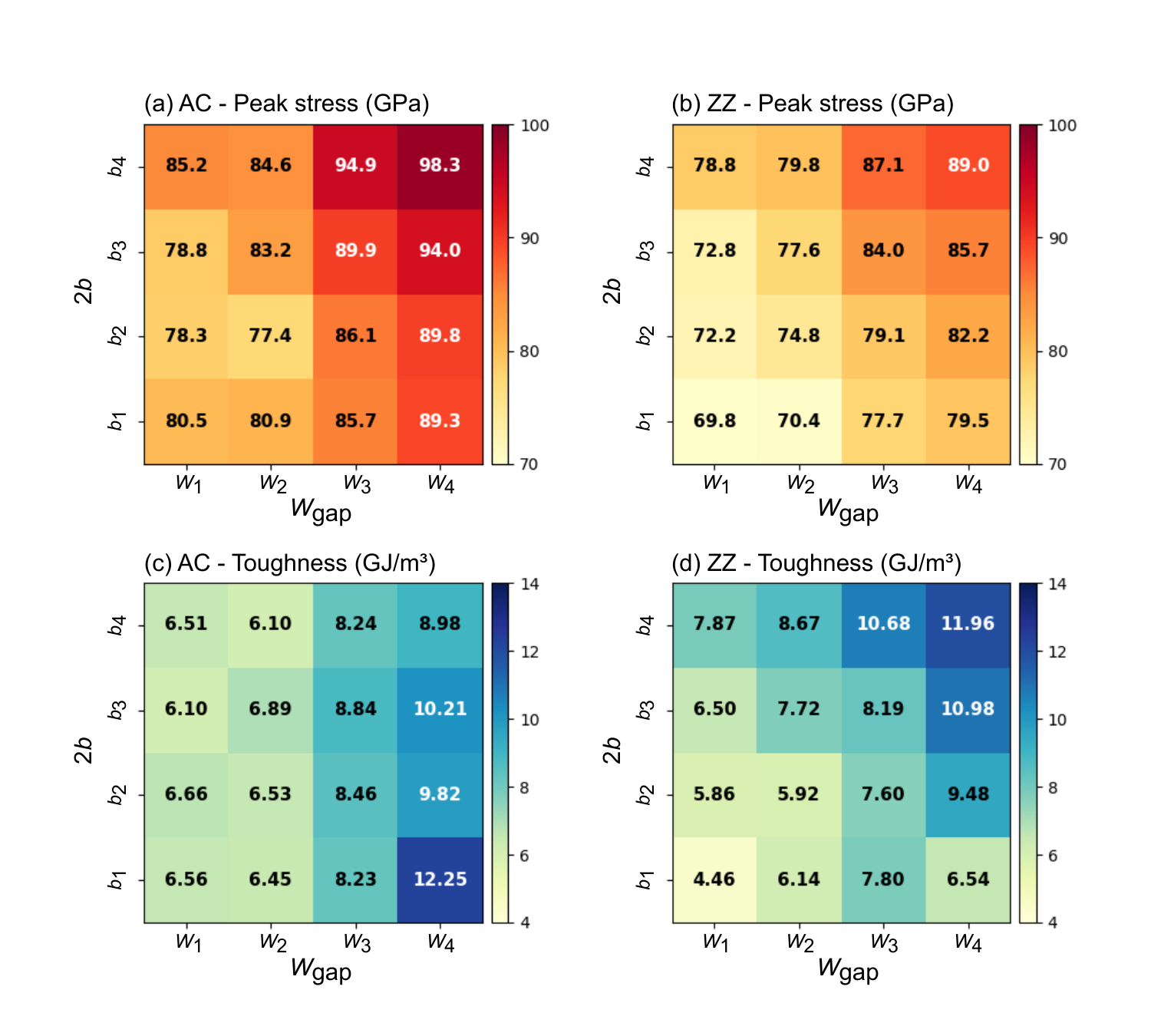}
    \caption{\label{fig:heatmap}Heat maps of (a,b) peak stress and (c,d) fracture toughness of DC graphene over the full $4\times4$ parameter matrix defined by crack width, $2b$, and inter-crack gap, $W_\text{gap}$, for AC and ZZ chiralities. The crack-width levels ($b_1$--$b_4$) and crack gap levels ($w_1$--$w_4$) are arranged in ascending order. For each mechanical property, identical colour scales are used for the AC and ZZ panels to enable direct comparison between chiralities. Fracture toughness is defined as the area under the stress--strain curve up to the fracture strain,~$\varepsilon_f$.}    
\end{figure}

Fig.~\ref{fig:heatmap} summarizes the dependence of peak stress and toughness on crack width, $2b$, and inter-crack gap, $W_\text{gap}$, over the investigated $4\times4$ parameter space. 
For both AC and ZZ graphene, peak stress generally increases with increasing $W_\text{gap}$; the larger inter-crack gap enhances structural strength. 
Toughness also generally increases with $W_\text{gap}$; however, it is more sensitive than peak stress to the combined effects of crack width and chirality. 
The highest toughness among the AC configurations occurs at the narrowest crack width ($2b\approx0.61$~nm) and the largest crack gap ($W_\text{gap}=3.56$~nm). In contrast, the highest toughness among the ZZ configurations occurs at the largest crack width ($2b\approx3.76$~nm) and the largest crack gap ($W_\text{gap}=3.69$~nm). Therefore, the crack-width condition promoting toughness enhancement is governed by lattice chirality within the investigated parameter range. 
Overall, these results show that the fracture response of DC graphene is governed by the combined effects of crack width, crack gap, and chirality, rather than by crack gap alone.

\begin{figure}[tbp]
    \centering
    \includegraphics[width=\linewidth]{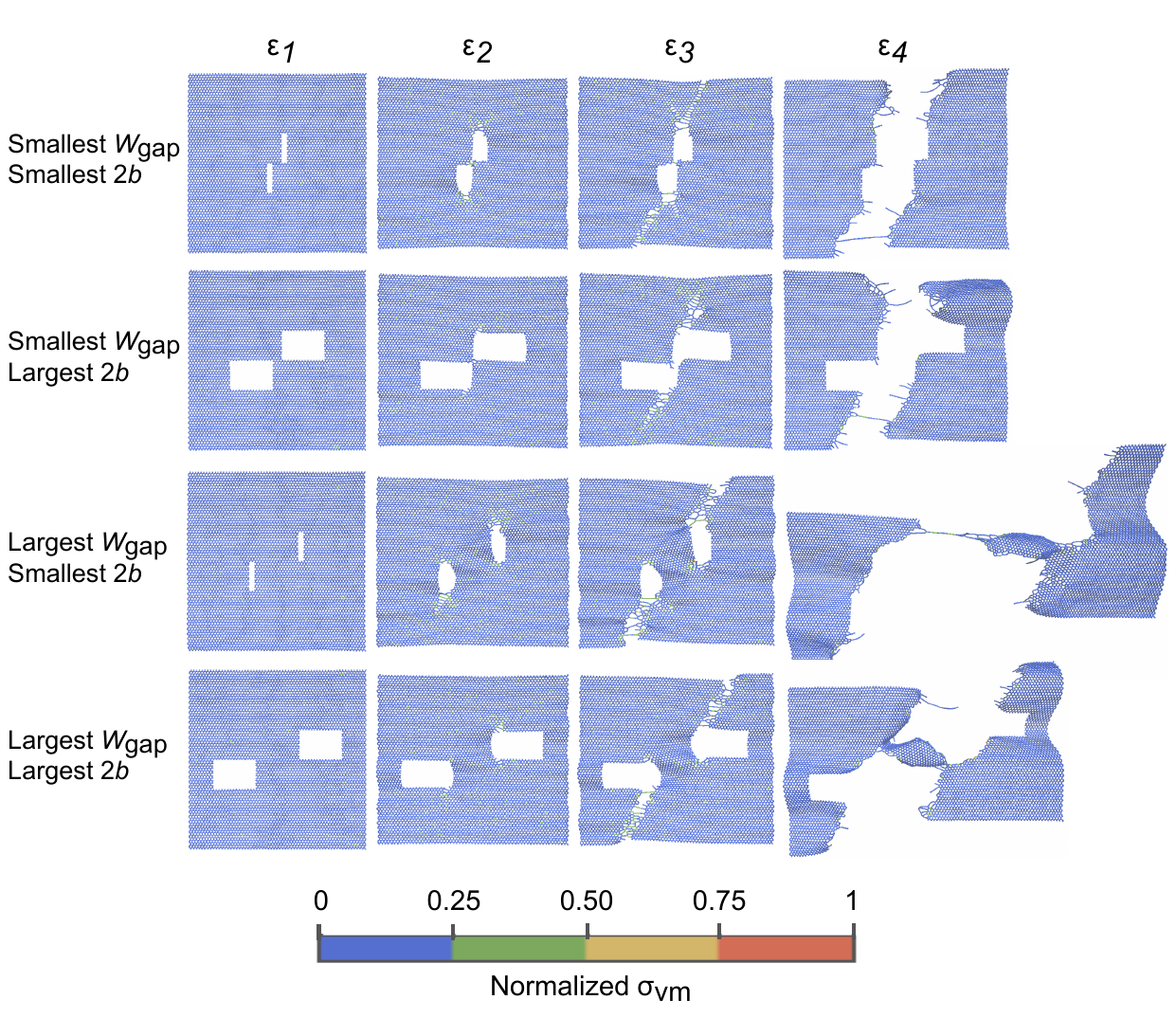}
    \caption{\label{fig:fracture-snapshot-AC}Snapshots of armchair
      (AC) graphene structures with dual cracks at different crack
      widths ($2b$) and inter-crack gaps ($W_\text{gap}$) during
      tensile deformation along the \textit{x}-direction. Rows
      correspond to increasing crack width and/or crack spacing, while
      columns represent different stages of deformation: initial
      configuration, near peak stress, post-peak deformation, and near
      final rupture. The color contours indicate the normalized von
      Mises stress distribution.  For the smallest $2b$ and
      $W_\text{gap}$, rapid crack coalescence occurs through the
      ligament between cracks, leading to brittle fracture. As $2b$
      and $W_\text{gap}$ increase, the ligament undergoes progressive
      deformation and rotation prior to rupture, resulting in delayed
      fracture and enhanced energy dissipation. The same normalized
      von Mises stress scale is used for all configurations to enable
      direct comparison of stress redistribution and crack
      interaction.}
\end{figure}

\begin{figure}[tbp]
    \centering
    \includegraphics[width=\linewidth]{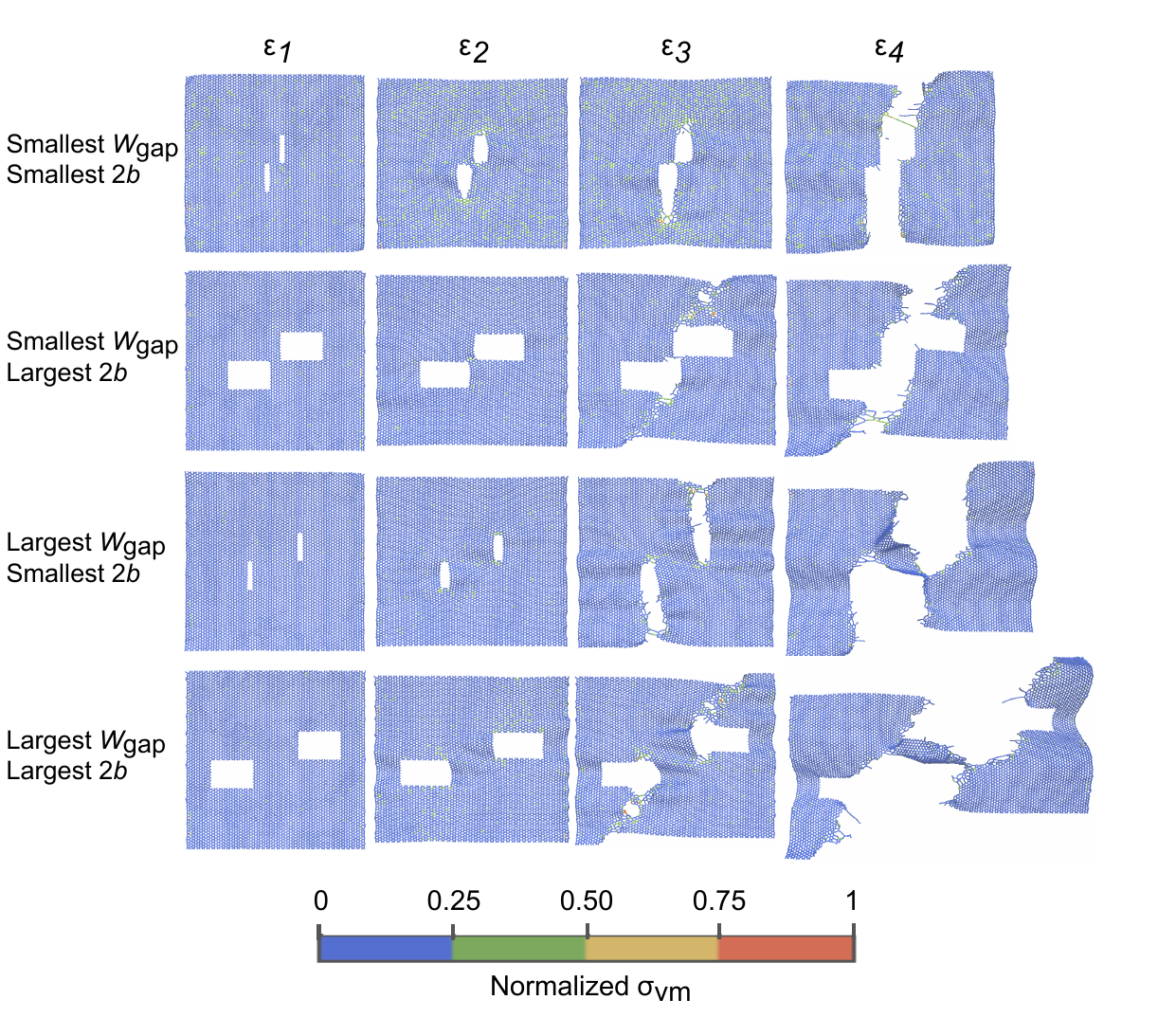}
    \caption{\label{fig:fracture-snapshot-zz}Snapshots of zigzag (ZZ)
      graphene structures with dual cracks at different crack widths
      (2b) and inter-crack gaps ($W_\text{gap}$) during tensile
      deformation along the \textit{x}-direction. Rows correspond to
      increasing crack width and/or crack spacing, while columns
      represent different stages of deformation: initial
      configuration, near peak stress, post-peak deformation, and near
      final rupture. The color contours indicate the normalized von
      Mises stress distribution. For small crack widths and crack
      gaps, fracture is dominated by crack coalescence accompanied by
      localized bond breaking between the inner crack tips. Increasing
      $2b$ and $W_\text{gap}$ promotes independent crack propagation
      and lever-like ligament deformation, which delays catastrophic
      rupture and increases toughness. Compared with AC graphene, ZZ
      graphene exhibits stronger localization of deformation and more
      pronounced anisotropic crack propagation behavior. The same
      normalized von Mises stress scale is used for all configurations
      to enable direct comparison of stress redistribution and crack
      interaction.}
\end{figure}

To clarify the fracture mechanisms underlying the geometry-dependent trends in Fig.~\ref{fig:heatmap}, atomistic fracture snapshots and normalized von Mises stress distributions are examined for representative limiting cases of crack width and inter-crack gap. Figures~\ref{fig:fracture-snapshot-AC} and~\ref{fig:fracture-snapshot-zz} show the fracture evolution of AC and ZZ graphene, respectively, for four selected configurations corresponding to the corner regions of the $2b$–$W_\text{gap}$ parameter space: smallest $W_\text{gap}$–smallest $2b$, largest $W_\text{gap}$–smallest $2b$, smallest $W_\text{gap}$–largest $2b$, and largest $W_\text{gap}$–largest $2b$. These representative cases provide mechanistic insight into the toughness distribution observed in the heat maps.
At the smallest $W_\text{gap}$, both AC and ZZ graphene show strong interaction between neighboring cracks. Stress localization develops in the ligament region between the inner crack tips, followed by rapid crack coalescence during tensile deformation. This localized coalescence limits post-peak deformation and is consistent with the lower-toughness regions observed for closely spaced cracks in Fig.~\ref{fig:heatmap}. 
At the largest $W_\text{gap}$, crack coalescence is delayed and the fracture process becomes more progressive. 
In AC graphene with the smallest $2b$, this delayed interaction is accompanied by extended deformation before final rupture, which is consistent with the high-toughness response of the narrow-crack AC configuration in Fig.~\ref{fig:heatmap}. In ZZ graphene, however, the highest toughness is observed for the largest $2b$ at large $W_\text{gap}$, where more pronounced ligament deformation and crack-path development occur prior to final failure.
Overall, the snapshot analysis indicates that the geometry-dependent toughness enhancement originates from a transition in fracture mode. Configurations with small $W_\text{gap}$ promote localized crack coalescence and premature failure, while those with large $W_\text{gap}$ delay coalescence and allow more distributed deformation before final rupture. This transition provides a mechanistic basis for the heat-map observation that toughness depends strongly on the combined effects of $2b$, $W_\text{gap}$, and chirality.

\section{Summary and Conclusion}\label{sec4}
In this study, molecular dynamics simulations are performed to investigate the coupled effects of crack width, $2b$, and inter--crack gap, $W_\text{gap}$, on the tensile fracture behavior of graphene containing two parallel cracks. Armchair (AC) and zigzag (ZZ) graphene sheets are examined under uniaxial tensile loading and compared with corresponding single-crack (SC) configurations having the same crack width and an equivalent total crack length. The mechanical response is evaluated in terms of stress--strain behavior, peak stress, fracture strain, and toughness, while atomistic fracture snapshots and normalized von Mises stress distributions are used to clarify the underlying crack interaction mechanisms.
The results show that crack interaction affects the energy absorption capacity more strongly than the peak tensile strength. 
The normalized toughness ratio, $E_\text{DC}/E_\text{SC}$, ranges from 0.79 to 2.02, compared with 0.98--1.26 for the normalized peak-stress ratio; the crack geometry primarily influences post-peak deformation and fracture progression. 
Configurations with small $W_\text{gap}$ tend to accelerate crack coalescence and reduce energy absorption, while those with large $W_\text{gap}$ promote delayed fracture and toughness enhancement.
These findings provide atomistic insight into geometry-controlled crack interaction and suggest a potential route for tailoring the fracture resistance of graphene-based nanostructures through controlled defect architecture.

\section{Methodology}\label{sec2}
\subsection{\label{sec2.1}Graphene structures and crack geometry}
The graphene and crack geometries are considered as shown in
Fig.~\ref{fig:geometry}.  The dimensions of the graphene structure are
$15\times15~\text{nm}^2$ for all configurations. In contrast to our
previous study~\cite{suyeong2026IJMS}, the inter-crack gap
($W_\text{gap}$) in the present work is defined as the edge-to-edge
distance between neighboring cracks rather than the center-to-center
distance, enabling a more consistent geometrical interpretation of
crack interaction and ligament deformation for different crack widths.
Both single-crack (SC) and dual-crack (DC) graphene structures are
prepared to investigate the coupled effects of crack width ($2b$) and
inter-crack spacing ($W_\text{gap}$) on the mechanical response and
fracture behavior of graphene.  For the SC configurations, a single
rectangular crack with length ($2a_0$) is introduced at the center of
the graphene sheet. For the DC configurations, two parallel cracks
with individual crack lengths ($2a_1$), satisfying ($2a_1 \approx
a_0$), are separated by $W_\text{gap}$. The crack width of each crack
is set to $2b$. Four crack widths ($2b$) and five crack gaps
($W_\text{gap}$) are considered, resulting in four SC cases and 16 DC
cases for each graphene chirality, namely armchair (AC) and zigzag
(ZZ), corresponding to 20 cases per chirality and 40 cases in total
(see \ref{Appendix:geometry}). The crack edges are hydrogen-passivated
to stabilize dangling bonds and suppress artificial edge effects
during tensile deformation. A magnified view of the atomic structure
surrounding the cracks is also presented, as shown in
Fig.~\ref{fig:geometry}, where carbon and hydrogen atoms are shown in
cyan and blue, respectively.

\subsection{Molecular dynamics simulations}
The prepared graphene structures, totaling 40 cases, are investigated
under uniaxial tensile loading along the $x$-direction. Molecular
dynamics (MD) simulations are performed using the LAMMPS
package~\cite{lammps2022}. Interatomic interactions between carbon and
hydrogen atoms are described using the reactive force field
(ReaxFF)~\cite{vanDuin2001}, which enables accurate modeling of bond
breaking and bond formation processes during fracture. The set of
parameters developed by Chenoweth, van Duin and
Goddard~\cite{Chenoweth2008JPCA} was used as in our previous
study~\cite{suyeong2026IJMS}.

Before tensile loading, each graphene structure is relaxed through a
multi-step energy minimization procedure using the conjugate gradient
algorithm.  The convergence criteria, set to 10$^{-9}$ (dimensionless)
for energy and 10$^{-9}$ (kcal/mol)/\AA \ for force, are chosen to
remove residual stresses and obtain a stable initial configuration.
After minimization, atomic velocities are assigned to the free atoms
according to a temperature of 300 K, and the structures are thermally
equilibrated for 200,000 timesteps using a Langevin thermostat with a
damping factor of 100 timesteps. The simulation timestep was set to
0.05~fs.

Uniaxial tensile loading is applied along the $x$-direction after
equilibration. Narrow boundary regions are defined at the left and
right edges of the graphene sheet. The left boundary atoms are fixed
in the $x$-direction, while the right boundary atoms are assigned a
constant velocity in the $x$-direction corresponding to an engineering
strain rate of $10^8~\text{s}^{-1}$. The transverse motion of the
boundary atoms is allowed, enabling lateral relaxation during
stretching. The tensile simulations were continued up to the rupture
of the structure. The system temperature is maintained at 300~K during
tensile deformation using a Langevin thermostat applied to the free
atoms.

During tensile loading, stress–strain data are recorded every 500
timesteps, and atomic configurations are exported periodically for
post-processing. The exported configurations are used to examine crack
propagation, crack coalescence, ligament deformation, and final
fracture mechanisms.

\subsection{Stress calculation and toughness}
The per-atom stresses computed by LAMMPS are expressed in units of
stress multiplied by volume. To convert to engineering stress, the
system volume must be determined. As in our previous
study~\cite{suyeong2026IJMS}, the Voronoi tessellation method is
employed to estimate atomic volumes in deformed graphene
configurations containing crack openings and out-of-plane undulations.
Since the Voronoi tessellation method yields an effective volume
larger than the conventional quasi-two-dimensional graphene volume,
the calculated volume is corrected using a scaling factor of
$c=6.067$, previously determined by comparing the Young’s moduli
obtained from the Voronoi-based and conventional volume definitions in
the linear elastic regime~\cite{suyeong2026IJMS}. The scaling factor
allows us to obtain the precise engineering stress as if it was
calculated from the total virial stress divided by the corrected
system volume.


The fracture strain, ($\varepsilon_f$), is defined as the strain at
which the tensile stress decreased below $1$~GPa during tensile
deformation. This criterion is adopted to provide a consistent
definition of final fracture for all crack configurations. The
toughness is evaluated as the area under the stress--strain curve up
to the fracture strain, representing the total mechanical energy
absorbed by the graphene sheet prior to final failure.

\appendix
\renewcommand{\thefigure}{\arabic{figure}} 
\setcounter{figure}{7}
\renewcommand{\thetable}{\arabic{table}}
\setcounter{table}{0}

\section{Geometric information of Graphene Structure\label{Appendix:geometry}}

As explained in Fig.~\ref{fig:geometry} in section~\ref{sec2.1}, a total of 40 cases are generated, and the geometric information is listed in Table~\ref{tab:geometry}. 
\begin{table*}[htbp]
\centering
\caption{Geometric parameters of all simulated single-crack (SC) and dual-crack (DC) graphene cases measured from atomic coordinates. $2a_0$: crack length of the SC configuration; $2a_1$: individual crack length in the DC configuration; $2b$: crack width; $W_\text{gap}$: inter-crack gap. Dashes (--) indicate parameters not applicable to the configuration. The cases are grouped by chirality and crack type, and then listed in ascending order of crack width $2b$ and inter-crack gap $W_\text{gap}$.}
\label{tab:geometry}
\begin{tabular}{clcccc|clcccc}
\hline
\multicolumn{6}{c|}{\textbf{AC chirality}} & \multicolumn{6}{c}{\textbf{ZZ chirality}} \\
\hline
No. & Type & $2a_0$ & $2a_1$ & $2b$ & $W_\text{gap}$ &
No. & Type & $2a_0$ & $2a_1$ & $2b$ & $W_\text{gap}$ \\
& & (nm) & (nm) & (nm) & (nm) &
& & (nm) & (nm) & (nm) & (nm) \\
\hline
1  & SC & 5.317 & -- & 0.614 & --    & 21 & SC & 5.158 & -- & 0.567 & -- \\
2  & SC & 5.317 & -- & 1.351 & --    & 22 & SC & 5.158 & -- & 1.418 & -- \\
3  & SC & 5.317 & -- & 2.579 & --    & 23 & SC & 5.158 & -- & 2.694 & -- \\
4  & SC & 5.317 & -- & 3.807 & --    & 24 & SC & 5.158 & -- & 3.970 & -- \\
\hline
5  & DC & -- & 2.765 & 0.614 & 0.614 & 25 & DC & -- & 2.702 & 0.567 & 0.709 \\
6  & DC & -- & 2.765 & 0.614 & 1.105 & 26 & DC & -- & 2.702 & 0.567 & 1.134 \\
7  & DC & -- & 2.765 & 0.614 & 2.333 & 27 & DC & -- & 2.702 & 0.567 & 2.411 \\
8  & DC & -- & 2.765 & 0.614 & 3.561 & 28 & DC & -- & 2.702 & 0.567 & 3.687 \\
\hline
9  & DC & -- & 2.765 & 1.351 & 0.614 & 29 & DC & -- & 2.702 & 1.276 & 0.709 \\
10 & DC & -- & 2.765 & 1.351 & 1.105 & 30 & DC & -- & 2.702 & 1.418 & 1.134 \\
11 & DC & -- & 2.765 & 1.351 & 2.333 & 31 & DC & -- & 2.702 & 1.276 & 2.411 \\
12 & DC & -- & 2.765 & 1.351 & 3.561 & 32 & DC & -- & 2.702 & 1.276 & 3.687 \\
\hline
13 & DC & -- & 2.765 & 2.579 & 0.614 & 33 & DC & -- & 2.702 & 2.482 & 0.709 \\
14 & DC & -- & 2.765 & 2.579 & 1.105 & 34 & DC & -- & 2.702 & 2.552 & 1.134 \\
15 & DC & -- & 2.765 & 2.579 & 2.333 & 35 & DC & -- & 2.702 & 2.552 & 2.411 \\
16 & DC & -- & 2.765 & 2.579 & 3.561 & 36 & DC & -- & 2.702 & 2.482 & 3.687 \\
\hline
17 & DC & -- & 2.765 & 3.807 & 0.614 & 37 & DC & -- & 2.702 & 3.758 & 0.709 \\
18 & DC & -- & 2.765 & 3.807 & 1.105 & 38 & DC & -- & 2.702 & 3.758 & 1.134 \\
19 & DC & -- & 2.765 & 3.807 & 2.333 & 39 & DC & -- & 2.702 & 3.758 & 2.411 \\
20 & DC & -- & 2.765 & 3.807 & 3.561 & 40 & DC & -- & 2.702 & 3.758 & 3.687 \\
\hline
\end{tabular}
\end{table*}

\section{Quantified Peak Stress and Fracture Strain Data Extracted from Stress–Strain Responses\label{Appendix:data}}
\begin{figure}[htbp]
    \centering
    \includegraphics[width=\linewidth]{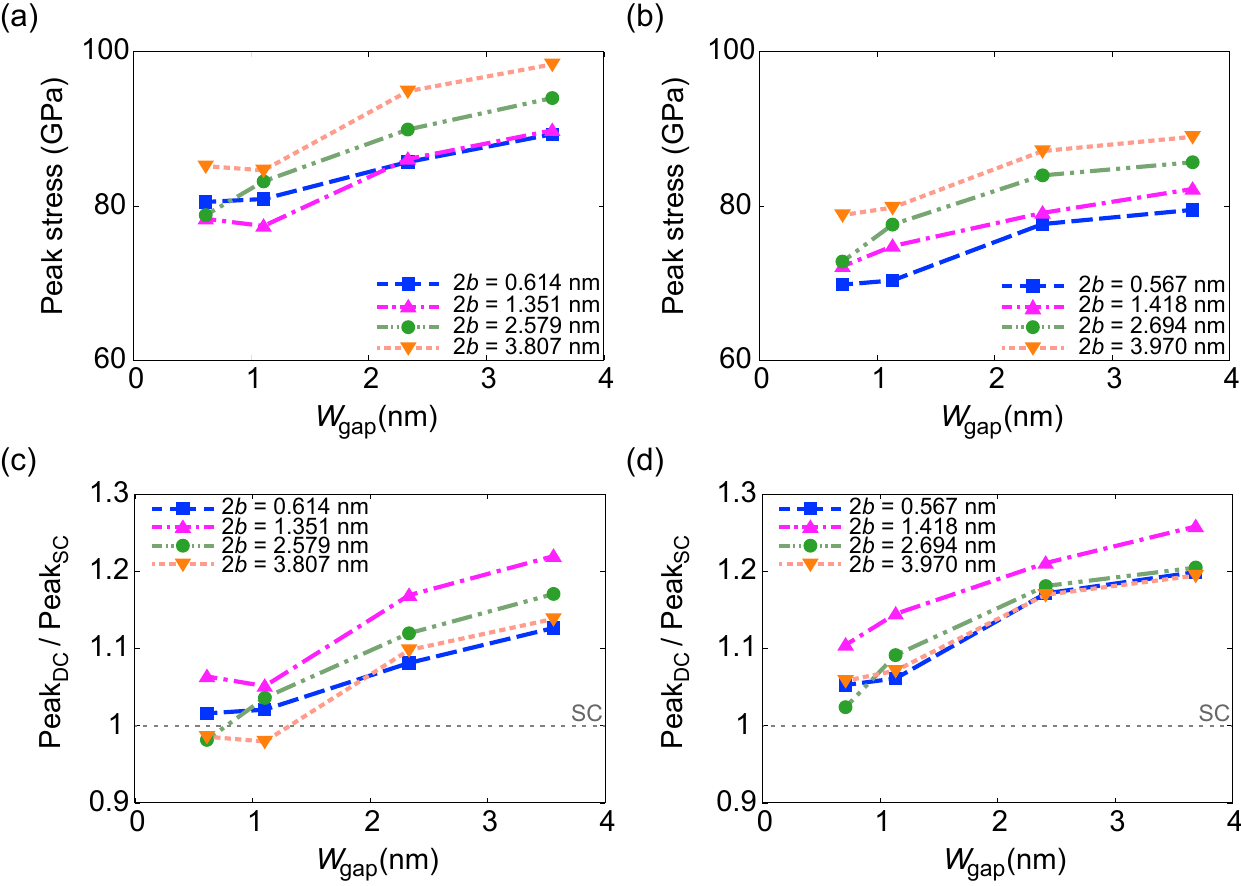}
    \caption{\label{fig:peak-vs-Wgap}Peak stress versus crack gap,
      $W_\text{gap}$, for graphene with two parallel cracks: (a)
      armchair (AC) and (b) zigzag (ZZ) chiralities; (c, d) peak
      stress ratio of the dual-crack configuration to the
      corresponding single-crack configuration with the same crack
      width, $2b$, as a function of $W_\text{gap}$, for AC and ZZ
      chiralities, respectively. The horizontal dashed line in (c) and
      (d) denotes unity, i.e., the peak stress of the equivalent
      single-crack (SC) configuration.  Each panel presents four
      curves corresponding to crack widths of $2b = 0.614, 1.351,
      2.579, \text{ and } 3.807$~nm for AC and $2b = 0.567, 1.418,
      2.694, \text{ and } 3.970$~nm for ZZ, respectively.}
\end{figure}

The peak stress and fracture strain values are extracted from the
stress–strain responses, as described in section~\ref{sec3}.
Figure~\ref{fig:peak-vs-Wgap} presents the peak stress of DC graphene
as a function of $W_\text{gap}$ for both AC and ZZ chiralities. For
all crack width ($2b$) groups and both chiralities, the peak stress
increased monotonically with increasing $W_\text{gap}$. The
sensitivity to $W_\text{gap}$ is most pronounced for intermediate
crack widths ($2b \approx 2.579$~nm), with a peak stress range of
approximately 15.2~GPa across the studied $W_\text{gap}$ values for
AC, while the narrowest crack ($2b \approx 0.61$~nm) showed the
smallest variation ($\sim$ 8.8~GPa). Across all conditions, AC
graphene consistently exhibited higher peak stress than ZZ graphene,
reflecting the intrinsic anisotropy of the graphene lattice.

The normalized peak-stress ratio, as shown in
Figs.~\ref{fig:peak-vs-Wgap}(c) and (d), is calculated by dividing the
peak stress of the dual-crack system by that of the corresponding SC
configuration with the same crack width and an equivalent total crack
length.  For AC graphene, three cases at small $W_\text{gap}$ values
($2b = 2.58$ and $3.81$~nm, $W_\text{gap} = 0.61$--$1.11$~nm) yield
values below unity, with a minimum of approximately 0.98. In contrast,
all ZZ configurations maintained values above unity, ranging
approximately from 1.02 to 1.26.

For both chiralities, the ratio increased consistently with $W_\text{gap}$, approaching a plateau at large separations where crack interaction becomes negligible; at $W_\text{gap} = 3.56$--$3.69$~nm, peak ratios reached 1.13--1.26, confirming that sufficiently separated dual cracks are intrinsically stronger than a single crack of twice the individual length.

\begin{figure}[htbp]
    \centering
    \includegraphics[width=\linewidth]{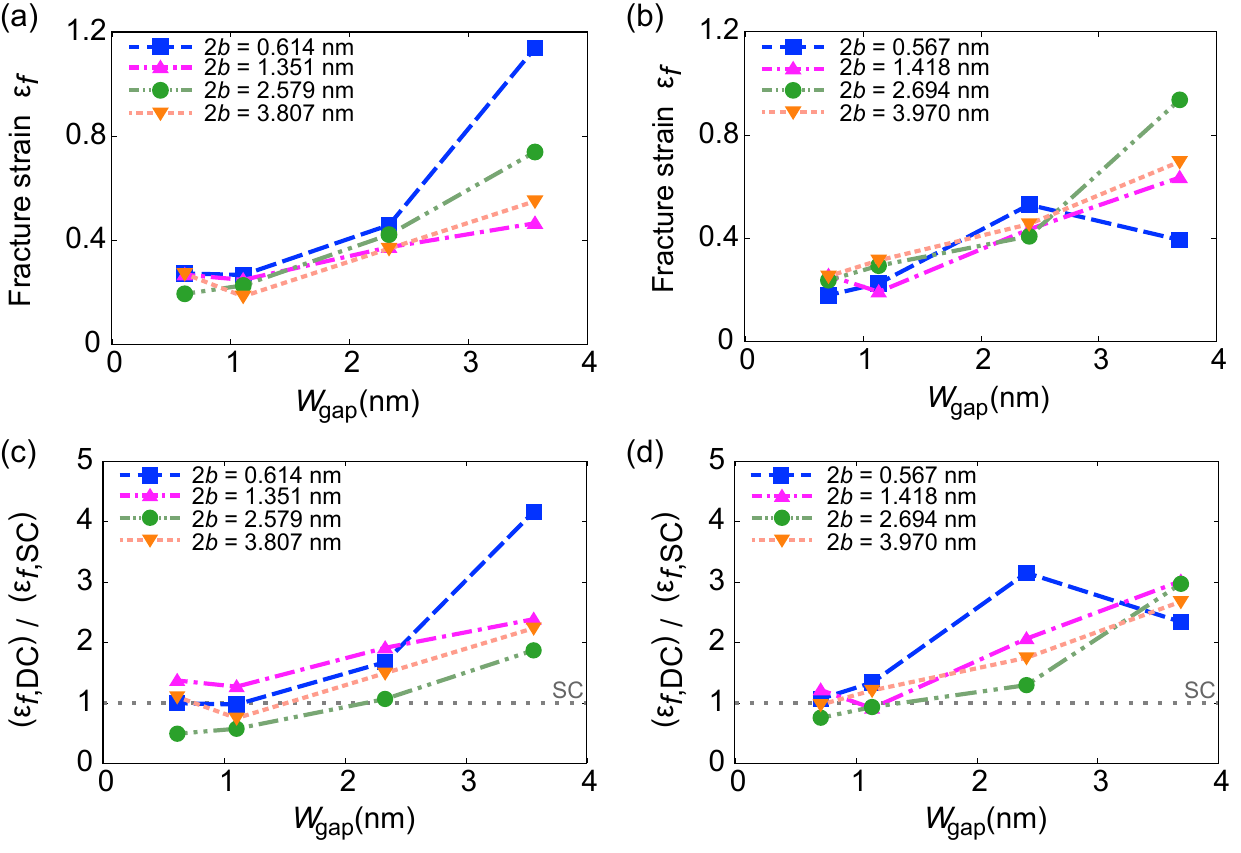}
    \caption{\label{fig:epsf-vs-Wgap}Fracture strain, $\varepsilon_f$, of dual-crack (DC) graphene sheets as a function of crack gap $W_\text{gap}$. Panels show results for (a) armchair (AC) and (b) zigzag (ZZ) chiralities. Panels (c) and (d) show the corresponding fracture strain ratio $\varepsilon_{f, \text{DC}}/\varepsilon_{f, \text{SC}}$ relative to the single-crack (SC) reference with the same crack width $2b$, for AC and ZZ, respectively. The horizontal dashed line at unity in (c) and (d) indicates the SC reference level. In particular, for the case ($2b = 2.579$~nm and $W_\text{gap}=2.333$~nm), graphene was fully broken before 1~GPa reached, and a lower-bound value is used for the fracture strain.}    
\end{figure}

Fig.~\ref{fig:epsf-vs-Wgap} presents the fracture strain ($\varepsilon_f$) of DC graphene versus $W_\text{gap}$ for both AC and ZZ chiralities. For both chiralities, $\varepsilon_f$ increases substantially with increasing $W_\text{gap}$, and this trend is consistently observed across all $2b$ groups. Narrow-crack configurations (small $2b$) exhibit the most pronounced enhancement in fracture strain at large $W_\text{gap}$, while wider-cracks (large $2b$) show comparatively moderate increases. 
The normalized fracture-strain ratio $(\varepsilon_{f,\text{DC}})/(\varepsilon_{f,\text{SC}})$ is as shown in Figs.~\ref{fig:epsf-vs-Wgap}(c) and (d). At small $W_\text{gap}$, several DC configurations yield ratios below unity, particularly for intermediate and wide cracks. For example, the AC configuration with $2b\approx 2.58$~nm at $W_\text{gap} = 0.61$~nm exhibits a minimum ratio of 0.49, and the corresponding ZZ configuration ($2b\approx 2.69$~nm) exhibits about 0.75. 
With increasing $W_\text{gap}$, the fracture-strain ratio increases for all crack-width groups, reaching values up to 4.16 for AC graphene ($2b \approx 0.61$~nm, $W_\text{gap} = 3.56$~nm) and 3.15 for ZZ graphene ($2b \approx 0.57$~nm, $W_\text{gap} = 2.41$~nm). 
Overall, fracture strain shows substantially greater variation across the parameter space than peak stress, with normalized ratios ranging from 0.49 to 4.16 compared with 0.98 to 1.26 for peak stress. This observation indicates that crack interaction influences the ductility of dual-crack graphene more strongly than the peak tensile strength.

\newpage

\section*{Acknowledgments}
The authors thank the Brazilian Coordination for the Improvement of Higher Education Personnel (CAPES) for covering the open access publication costs. Computational resources were provided by the Coaraci Supercomputer for computer time (Fapesp grant \#2019/17874-0) and the Center for Computing in Engineering and Sciences at Unicamp (Fapesp grant \#2013/08293-7). AFF thanks the computing resources and assistance of the John David Rogers Computing Center at the Institute of Physics Gleb Wataghin of the University of
Campinas. 

\section*{Funding}

This work was supported by the Pukyong National University Research
Fund in 2025(202516520001). AFF acknowledges the National Council for
Scientific and Technological Development (CNPq) -- Brazil for the
financial support and fellowship (grant number 302009/2025-6) and the
São Paulo Research Foundation (FAPESP) (grant number 2024/14403-4).

\section*{Data availability}

The authors declare that raw data supporting the findings of
this study are available upon request.

\section*{Materials availability}

Not applicable.

\section*{Code availability}

Simulations codes are freely available at \url{https://www.lammps.org/}.

\section*{Declarations}

\noindent {\bf Conflict of interest:} The authors declare no competing financial
interest.

\bibliographystyle{elsarticle-num}
\bibliography{main}

@article{suyeong2026IJMS,
title = {Abnormal crack coalescence and ductility in graphene},
journal = {International Journal of Mechanical Sciences},
volume = {309},
pages = {111025},
year = {2026},
issn = {0020-7403},
doi = {https://doi.org/10.1016/j.ijmecsci.2025.111025},
author = {Suyeong Jin and Jung-Wuk Hong and Chiara Daraio and Alexandre F. Fonseca}
}

@article{han2017MSER,
title = {Graphene-based flexible electronic devices},
journal = {Materials Science and Engineering: R: Reports},
volume = {118},
pages = {1-43},
year = {2017},
issn = {0927-796X},
doi = {https://doi.org/10.1016/j.mser.2017.05.001},
author = {Tae-Hee Han and Hobeom Kim and Sung-Joo Kwon and Tae-Woo Lee}
}

@article{quesnel2015graphene,
  title={Graphene-based technologies for energy applications, challenges and perspectives},
  author={Quesnel, Etienne and Roux, Fr{\'e}d{\'e}ric and Emieux, Fabrice and Faucherand, Pascal and Kymakis, Emmanuel and Volonakis, George and Giustino, Feliciano and Mart{\'\i}n-Garc{\'\i}a, Beatriz and Moreels, Iwan and G{\"u}rsel, Selmiye Alkan and others},
  journal={2D Materials},
  volume={2},
  number={3},
  pages={030204},
doi = {https://doi.org/10.1088/2053-1583/2/3/030204},
  year={2015},
  publisher={IOP Publishing}
}

@article{dhinakaran2020review,
  title={Review on exploration of graphene in diverse applications and its future horizon},
  author={Dhinakaran, V and Lavanya, M and Vigneswari, K and Ravichandran, M and Vijayakumar, MD},
  journal={Materials Today: Proceedings},
  volume={27},
  pages={824--828},
doi = {https://doi.org/10.1016/j.matpr.2019.12.369},
  year={2020},
  publisher={Elsevier}
}

@Article{Lin2009NatComm,
author={Lin, Li and Zhang, Jincan and Su, Haisheng and Li, Jiayu and Sun, Luzhao and Wang, Zihao and Xu, Fan and Liu, Chang and Lopatin, Sergei and Zhu, Yihan and Jia, Kaicheng and Chen, Shulin and Rui, Dingran and Sun, Jingyu and Xue, Ruiwen and Gao, Peng and Kang, Ning and Han, Yu and Xu, H. Q. and Cao, Yang and Novoselov, K. S. and Tian, Zhongqun and Ren, Bin and Peng, Hailin and Liu, Zhongfan},
title={Towards super-clean graphene},
journal={Nature Communications},
year={2019},
month={Apr},
day={23},
volume={10},
number={1},
pages={1912},
issn={2041-1723},
doi={https://doi.org/10.1038/s41467-019-09565-4},
}

@Article{Boggild2023NatComm,
author={B{\o}ggild, Peter},
title={Research on scalable graphene faces a reproducibility gap},
journal={Nature Communications},
year={2023},
month={Feb},
day={28},
volume={14},
number={1},
pages={1126},
issn={2041-1723},
doi={https://doi.org/10.1038/s41467-023-36891-5},
}

@Article{Pham2024,
  author    = {Pham, Phuong V. and Mai, The-Hung and Dash, Saroj P. and Biju, Vasudevanpillai and Chueh, Yu-Lun and Jariwala, Deep and Tung, Vincent},
  title     = {Transfer of 2D Films: From Imperfection to Perfection},
  journal   = {ACS Nano},
  year      = {2024},
  volume    = {18},
  number    = {23},
  pages     = {14841--14876},
  month     = jun,
  booktitle = {ACS Nano},
  doi       = {https://doi.org/10.1021/acsnano.4c00590},
  issn      = {1936-0851},
  publisher = {American Chemical Society},
}

@Article{Zandiatashbar2014NatComm,
author={Zandiatashbar, Ardavan
and Lee, Gwan-Hyoung
and An, Sung Joo
and Lee, Sunwoo
and Mathew, Nithin
and Terrones, Mauricio
and Hayashi, Takuya
and Picu, Catalin R.
and Hone, James
and Koratkar, Nikhil},
title={Effect of defects on the intrinsic strength and stiffness of graphene},
journal={Nature Communications},
year={2014},
month={Jan},
day={24},
volume={5},
number={1},
pages={3186},
issn={2041-1723},
doi={https://doi.org/10.1038/ncomms4186}
}

@article{Mahesh2022RSCAdv,
author ="Bhatt, Mahesh Datt and Kim, Heeju and Kim, Gunn",
title  ="Various defects in graphene: a review",
journal  ="RSC Adv.",
year  ="2022",
volume  ="12",
issue  ="33",
pages  ="21520-21547",
publisher  ="The Royal Society of Chemistry",
doi  ="https://doi.org/10.1039/D2RA01436J"
}

@article{Dongbo2024EFM,
title = {Fracture behavior of graphene with intrinsic defects and externally introduced defects},
journal = {Engineering Fracture Mechanics},
volume = {303},
pages = {110130},
year = {2024},
issn = {0013-7944},
doi = {https://doi.org/10.1016/j.engfracmech.2024.110130},
author = {Dongbo Li and Yihang Zhang and Jiapeng Guo and Jing Zhu and Qinlong Liu and Na Ni and Jiaqi Yan}
}

@article{Liu2025JPCS,
title = {Effect and regulation of pore defects on mechanical properties of graphene},
journal = {Journal of Physics and Chemistry of Solids},
volume = {206},
pages = {112870},
year = {2025},
issn = {0022-3697},
doi = {https://doi.org/10.1016/j.jpcs.2025.112870},
author = {Ying Liu and Jian-Gang Guo and Zhi-Na Zhao}
}

@article{DewaCarbon2017, 
title = {Atomistic simulations of nanoscale crack-vacancy interaction in graphene},
journal = {Carbon},
volume = {125},
pages = {113-131},
year = {2017},
issn = {0008-6223},
doi = {https://doi.org/10.1016/j.carbon.2017.09.015},
author = {M.A.N. Dewapriya and S.A. Meguid}
}

@article{DewaEngFracMech2018, 
title = {Atomistic modelling of crack-inclusion interaction in graphene},
journal = {Engineering Fracture Mechanics},
volume = {195},
pages = {92-103},
year = {2018},
issn = {0013-7944},
doi = {https://doi.org/10.1016/j.engfracmech.2018.04.003},
author = {M.A.N. Dewapriya and S.A. Meguid and R.K.N.D. Rajapakse}
}

@article{DewaCMS2018, 
title = {Tailoring fracture strength of graphene},
journal = {Computational Materials Science},
volume = {141},
pages = {114-121},
year = {2018},
issn = {0927-0256},
doi = {https://doi.org/10.1016/j.commatsci.2017.09.005},
author = {M.A.N. Dewapriya and S.A. Meguid}
}

@article{YaoEngFracMech2019,
title = {Finite element analysis and molecular dynamics simulations of nanoscale crack-hole interactions in chiral graphene nanoribbons},
journal = {Engineering Fracture Mechanics},
volume = {218},
pages = {106571},
year = {2019},
issn = {0013-7944},
doi = {https://doi.org/10.1016/j.engfracmech.2019.106571},
author = {Jinchun Yao and Yuxuan Xia and Shuhong Dong and Peishi Yu and Junhua Zhao}
}

@article{BrodnikPRL2021,
  title = {Fracture Diodes: Directional Asymmetry of Fracture Toughness},
  author = {Brodnik, N. R. and Brach, S. and Long, C. M. and Ravichandran, G. and Bourdin, B. and Faber, K. T. and Bhattacharya, K.},
  journal = {Phys. Rev. Lett.},
  volume = {126},
  issue = {2},
  pages = {025503},
  numpages = {5},
  year = {2021},
  month = {Jan},
  publisher = {American Physical Society},
  doi = {https://doi.org/10.1103/PhysRevLett.126.025503},
}

@article{felixPCCP2022,
author ="Felix, Levi C. and Galvao, Douglas S.",
title  ="Guided fractures in graphene mechanical diode-like structures",
journal  ="Phys. Chem. Chem. Phys.",
year  ="2022",
volume  ="24",
issue  ="22",
pages  ="13905-13910",
publisher  ="The Royal Society of Chemistry",
doi  ="https://doi.org/10.1039/D2CP01207C"}

@article{Hu2015JAP,
    author = {Hu, Lin and Wyant, Spencer and Muniz, Andre R. and Ramasubramaniam, Ashwin and Maroudas, Dimitrios},
    title = {Mechanical behavior and fracture of graphene nanomeshes},
    journal = {Journal of Applied Physics},
    volume = {117},
    number = {2},
    pages = {024302},
    year = {2015},
    month = {01},
    issn = {0021-8979},
    doi = {https://doi.org/10.1063/1.4905583}
}

@article{Chen2020ACSANM,
author = {Chen, Mengxi and Christmann, Augusto M. and Muniz, Andre R. and Ramasubramaniam, Ashwin and Maroudas, Dimitrios},
title = {Molecular-Dynamics Analysis of Nanoindentation of Graphene Nanomeshes: Implications for 2D Mechanical Metamaterials},
journal = {ACS Applied Nano Materials},
volume = {3},
number = {4},
pages = {3613-3624},
year = {2020},
doi = {https://doi.org/10.1021/acsanm.0c00327}
}

@article{John2014Carbon,
title = {Orientation dependence of the fracture behavior of graphene},
journal = {Carbon},
volume = {66},
pages = {619-628},
year = {2014},
issn = {0008-6223},
doi = {https://doi.org/10.1016/j.carbon.2013.09.051},
author = {Young I. Jhon and Young Min Jhon and Geun Y. Yeom and Myung S. Jhon}
}

@article{Fujihara2015ACSNano, 
author = {Fujihara, Miho and Inoue, Ryosuke and Kurita, Rei and Taniuchi, Toshiyuki and Motoyui, Yoshihito and Shin, Shik and Komori, Fumio and Maniwa, Yutaka and Shinohara, Hisanori and Miyata, Yasumitsu},
title = {Selective Formation of Zigzag Edges in Graphene Cracks},
journal = {ACS Nano},
volume = {9},
number = {9},
pages = {9027-9033},
year = {2015},
doi = {https://doi.org/10.1021/acsnano.5b03079}
}

@Article{Zhang2012,
  author    = {Zhang, Teng and Li, Xiaoyan and Kadkhodaei, Sara and Gao, Huajian},
  title     = {Flaw Insensitive Fracture in Nanocrystalline Graphene},
  journal   = {Nano Lett.},
  year      = {2012},
  volume    = {12},
  number    = {9},
  pages     = {4605--4610},
  month     = sep,
  booktitle = {Nano Letters},
  doi       = {https://doi.org/10.1021/nl301908b},
  issn      = {1530-6984},
  publisher = {American Chemical Society},
}

@Article{Meng2015,
  author    = {Meng, Fanchao and Chen, Cheng and Song, Jun},
  title     = {Dislocation Shielding of a Nanocrack in Graphene: Atomistic Simulations and Continuum Modeling},
  journal   = {J. Phys. Chem. Lett.},
  year      = {2015},
  volume    = {6},
  number    = {20},
  pages     = {4038--4042},
  month     = oct,
  booktitle = {The Journal of Physical Chemistry Letters},
  doi       = {https://doi.org/10.1021/acs.jpclett.5b01815},
  publisher = {American Chemical Society},
}

@Article{Lopez-Polin2015,
  author    = {López-Polín, Guillermo and Gómez-Herrero, Julio and Gómez-Navarro, Cristina},
  title     = {Confining Crack Propagation in Defective Graphene},
  journal   = {Nano Lett.},
  year      = {2015},
  volume    = {15},
  number    = {3},
  pages     = {2050--2054},
  month     = mar,
  booktitle = {Nano Letters},
  doi       = {https://doi.org/10.1021/nl504936q},
  issn      = {1530-6984},
  publisher = {American Chemical Society},
}

@article{vanDuin2001, 
author = {van Duin, Adri C. T. and Dasgupta, Siddharth and Lorant, Francois and Goddard, William A.},
title = {ReaxFF: A Reactive Force Field for Hydrocarbons},
journal = {The Journal of Physical Chemistry A},
volume = {105},
number = {41},
pages = {9396-9409},
year = {2001},
doi = {https://doi.org/10.1021/jp004368u},
}

@article{lammps2022,
title = {LAMMPS - a flexible simulation tool for particle-based materials modeling at the atomic, meso, and continuum scales},
journal = {Computer Physics Communications},
volume = {271},
pages = {108171},
year = {2022},
issn = {0010-4655},
doi = {https://doi.org/10.1016/j.cpc.2021.108171},
author = {Aidan P. Thompson and H. Metin Aktulga and Richard Berger and Dan S. Bolintineanu and W. Michael Brown and Paul S. Crozier and Pieter J. {in 't Veld} and Axel Kohlmeyer and Stan G. Moore and Trung Dac Nguyen and Ray Shan and Mark J. Stevens and Julien Tranchida and Christian Trott and Steven J. Plimpton}
}

@article{Chenoweth2008JPCA,
author = {Chenoweth, Kimberly and van Duin, Adri C. T. and Goddard, William A.},
title = {ReaxFF Reactive Force Field for Molecular Dynamics Simulations of Hydrocarbon Oxidation},
journal = {The Journal of Physical Chemistry A},
volume = {112},
number = {5},
pages = {1040-1053},
year = {2008},
doi = {https://doi.org/10.1021/jp709896w},
}

\section*{Author contributions}

Suyeong Jin: Conceptualization, Visualization, Methodology,
Investigation, Formal analysis, Data curation, Writing – original draft.
Jung-Wuk Hong: Writing – review \& editing, Validation, Supervision. 
Alexandre F. Fonseca:
Conceptualization, Methodology, Data curation, Writing – review \& editing, Validation, Supervision.

\end{document}